\documentclass[prb,twocolumn,amsmath,amssymb]{revtex4}
\usepackage{latexsym,amsmath,graphics,graphicx}

\newcommand{\AF}{\mathrm{AF}}
\newcommand{\FM}{\mathrm{FM}}
\newcommand{\Ferri}{\mathrm{Ferri}}

\begin{document}
\title{Changing the Magnetic Configurations of  Nanoclusters Atom-by-Atom}

\author{Samir Lounis}\email{s.lounis@fz-juelich.de}
\author{Phivos Mavropoulos}
\author{Rudolf Zeller}
\author{Peter H.~Dederichs}
\author{Stefan Bl\"ugel}
\affiliation{Institut f\"ur
Festk\"orperforschung, Forschungszentrum J\"ulich, D-52425 J\"ulich,
Germany}

\date{\today}

\begin{abstract}
The Korringa-Kohn-Rostoker Green (KKR) function method for
non-collinear magnetic structures was applied on Mn and Cr ad-clusters
deposited on the Ni(111) surface. By considering various dimers,
trimers and tetramers, a large amount of collinear and non-collinear
magnetic structures is obtained.  Typically all compact clusters have
very small total moments, while the more open structures exhibit
sizeable total moments, which is a result of the complex frustration
mechanism in these systems. Thus, as the motion of a single adatom
changes the cluster structure from compact to open and vice versa,
this can be considered as a magnetic switch, which via the local
exchange field of the adatom allows to switch the cluster moment on
and off, and which might be useful for future nanosize information
storage.

\end{abstract}
\maketitle

\section{Introduction}
Future technologies will be based on the magnetic properties of
nanostructures. Such magnetic structures can be composed of magnetic
atoms in precise arrangements. The magnetic properties of each atom
can be profoundly influenced by its local environment. 
Recently, Gambardella {\it et al.}\cite{gambardella} showed 
that Co adatoms on
Pt(111) have giant magnetic anisotropy energy (MAE) which may open the
way to very high data storage densities.  Indeed, clusters have the
potential of increasing the density in information storage. One may
envision that future magnetic hard discs with information carried by
magnetic clusters, will have a storage density two orders of magnitude
larger than those used today.  Here we show, for elements without high
MAE, {\it e.g.}, adatoms with half-filled 3$d$ shells (Cr and Mn),
that the magnetic properties due to the competition of
antiferromagnetic (AF) exchange coupling may be used to switch the
moment configuration in small ad-clusters.  This may help reaching an
important goal for magnet's technological relevance {\it e.g.} the
ability of a magnet to store bits
of 0 and 1 and being stable against thermal fluctuations. The bit 0
can be considered when no (or very small) magnetic moment is measured
while high magnetic moment of the cluster can be considered as 1.
In fact, five years ago, Jamneala {\it et al.}\cite{jamneala}
investigated by STM Cr trimers deposited on Cu(111). They show that
moving a Cr adatom of a compact trimer leads to a switching from the
Kondo state to a magnetic one.

Magnetic excitations may degrade the performance of high-density
memories. Indeed, Heinrich {\it et al.}\cite{heinrich} could elucidate
the spin-flip of individual magnetic atoms that are dispersed on a
non-magnetic matrix using scanning tunneling microscopy (STM).  In the
present paper, we propose to use the exchange field created by an
adatom to switch the magnetic configuration of an ad-cluster from
one magnetic state to another one.

 Recently, there 
has been increasing interest in investigating non-collinear 
nanostructures on ferromagnetic\cite{lounis,lounis2,robles} 
or non-magnetic 
surfaces\cite{bergman,stocks,gotsis,demangeat,costa}. Here, 
we choose as a 
substrate a ferromagnetic (FM) $fcc$-Ni surface which
provides a magnetic coupling between the adatoms 
and the substrate atoms. The Ni(111) surface was chosen,
in which the surface geometry is triangular, meaning, in terms of
magnetic coupling\cite{bergman}, that a compact trimer with
antiferromagnetic interactions sitting on the surface suffer
necessarily magnetic frustration. This leads to the well-known
non-collinear Neel states being characterized by 120$^{\circ}$ angles
between the moments. Hence, we face in such a system an interplay
between the non-collinear coupling tendencies arizing from the
interaction among the adatoms in the cluster and the collinear
tendencies arizing from the additional coupling to the substrate
atoms: this is very different to the Ni(001) surface where the
frustration arizes from the competition between the coupling in the
cluster and with the substrate\cite{lounis}.

\section{Calculational Method}

Our calculations are based on the Local Spin Density Approximation
(LSDA) of density functional theory with the parametrization of Vosko
{\it et al.}\cite{vosko}. The full non-spherical potential was used,
taking into account the correct description of the Wigner--Seitz
atomic cells.\cite{Stefanou90,papanikolaou} Angular momenta up to
$l_{\mathrm{max}} = 3$ were included in the expansion of the Green
functions and up to $2l_{\mathrm{max}} = 6$ in the charge density
expansion. Relativistic effects were described in the scalar
relativistic approximation.

First, the surface Green functions are determined by the screened KKR
method\cite{SKKR} for the (111) surface of Ni which serves as the
reference system. The LSDA equilibrium lattice parameter of Ni was
used (6.46~{a.u.}  $\approx$~3.42~{\AA}) and the magnetic moment at
the surface is 0.63 $\mu_B$. To describe the Cr and Mn adatoms on the
surface we consider a cluster of perturbed potentials with a size of
48 perturbed sites for all kind of ad-clusters considered.  We
consider the adatoms at the unrelaxed hollow position in the first
vacuum layer.  We allow for the relaxation of the magnetic moment
directions with respect to the direction of the substrate
magnetization\cite{lounis}.
\section{Theoretical Background}
Since the collinear magnetic state represents a self-consistent
solution of the Kohn-Sham equations, total energy calculations are
necessary to check whether the non-collinear solution represents a
true energy minimum or only a local minimum, with the collinear state
representing the total minimum. This proved to be important, e.g., for
Cr dimers on Ni(001)~\cite{lounis}, where the collinear solution was
found to have a lower total energy.

The driving mechanism for non-collinear magnetism in small
transition metal clusters is a competition of antiferromagnetic (or
antiferromagnetic and ferromagnetic) interactions. In other cases,
such as $f$-element systems\cite{nordstrom} or ad-layers and 
chains\cite{dzialoshinskii,fischer}, the spin-orbit interaction can also
be of significance but in transition metals usually 
frustration is the dominating 
effect. In the case of magnetically non-collinear
transition metal clusters adsorbed on ferromagnetic surfaces, it is
helpful for the interpretation of the results to distinguish between
three factors contributing to the solution: (i) the magnetic 
interaction of the
cluster adatoms with the substrate, (ii) the magnetic 
pair interaction among
the atoms in the cluster, and (iii) the geometry of the cluster. This
separation is meaningful because the first-neighbours exchange
interaction is energetically dominant compared to second, third,
etc.~neighbours, and because in different cluster sizes or shapes the
type of pair interaction (ferro- or antiferromagnetic) does not change
qualitatively. Quantitatively, however, this is only an approximation,
and effects beyond this are included in the self-consistent solution.
In view of the above, we proceed by first presenting results for the
single adatoms, then the dimers, and then trimers and tetramers of
various shapes. We expect that Cr and Mn clusters are candidates for
non-collinear magnetism, because the Cr-Cr and Mn-Mn first-neighbors
pair interactions are antiferromagnetic.

\section{Single Adatoms and Dimers}
Investigating the magnetism of ad-clusters starts by understanding the magnetism of 
adatoms and ad-dimers.

{\it ---Single Adatoms.} Our calculations show that the single
Cr adatom is AF coupled to the surface with an increase of the
magnetic moments ($M_{AF}$ = 3.77 $\mu_B$ and $M_{FM}$ = 3.70 $\mu_B$)
compared to the results obtained for Ni(001)\cite{lounis} ($M_{AF}$ =
3.48 $\mu_B$ and $M_{FM}$ = 3.35 $\mu_B$).  This increase arizes from
the weaker hybridization of the $3d$ wavefunctions with the
substrate---the adatom has three neighbours on the (111) surface and
four on the (001).  The calculated energy difference between the FM
and AF configurations is high so that the AF configuration is stable
at room temperature ($\Delta E_{\AF-\FM}=-93.54$~meV, corresponding to
1085~K). Also our results for the Mn adatom on Ni(111) are similar to
what we found on Ni(001). The single Mn adatom prefers to couple
ferromagnetically to the substrate. The energy difference between the
two possible magnetic configurations is $\Delta E_{\AF-\FM}=208$~meV. 
For the (001) surfaces the energy
differences are both for Cr and Mn larger, since they roughly scale
with the coordination number ($\Delta
E_{\AF-\FM}^{\mathrm{Cr}}=-134$~meV, $\Delta
E_{\AF-\FM}^{\mathrm{Mn}}=252$~meV). The
magnetic moments of Mn are high and reach a value of 4.17 $\mu_B$ for
the FM configuration and 4.25 $\mu_B$ for the AF configuration. The
moments are higher than for the Mn adatoms on Ni(001) ($M_{AF}$ = 4.09
$\mu_B$ and $M_{FM}$ = 3.92 $\mu_B$), again due to the lower
coordination and hybridization of the $3d$ levels. This type of
coupling to the substrate (AF for Cr and FM for Mn) can be understood
in terms of the $d$-$d$ hybridization of the adatom wavefunctions with
the ones of the substrate, and is described, e.g., in
Ref.~\onlinecite{lounis}.

\begin{figure}
{(a)}
\includegraphics*[width=0.4\linewidth]{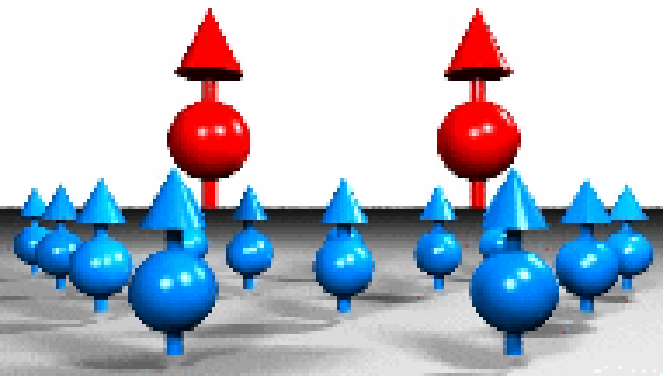}
\hspace{-0.2cm}
{(b)}
\includegraphics*[width=0.4\linewidth]{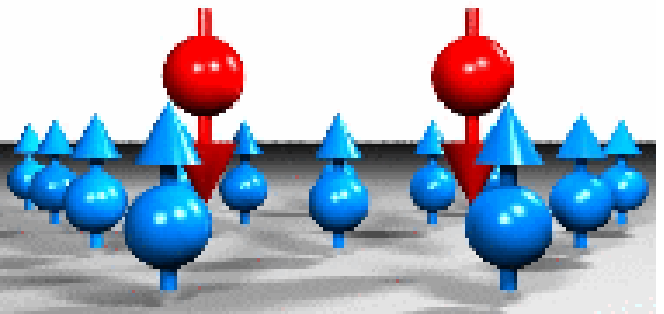}
\\
{(c)}
\includegraphics*[width=0.4\linewidth]{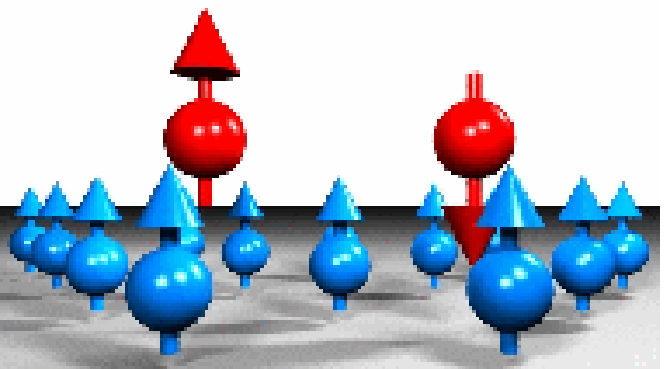}
\hspace{-0.2cm}
{(d)}
\includegraphics*[width=0.4\linewidth]{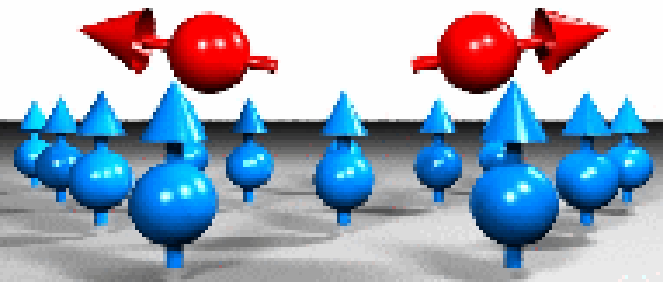}
\caption{Different magnetic configurations of Mn Dimer on Ni(001). The configurations 
correspond to FM in (a), AF in (b),  Ferri in (c) which is the ground state, and 
the non-collinear additional local minimum in (d). See text for the discussion.
}
\label{dimer-Ni111}
\end{figure}

{\it ---Dimers.} For the compact dimers, three collinear
configurations are possible: ferromagnetic (FM) 
(see fig.~\ref{dimer-Ni111}(a)), with the moments of
both atoms parallel to the substrate moments 
(fig.~\ref{dimer-Ni111}(b)), antiferromagnetic (AF),
with the moments of both atoms antiparallel to the substrate moments,
and ferrimagnetic (Ferri) (see fig.~\ref{dimer-Ni111}(c)), 
where the magnetic moment of one of the
dimer atoms is parallel to the moments of the substrate, while the
other one is antiparallel. Since the direct 
exchange in a Cr pair (or a Mn
pair) is antiferromagnetic~(for an explanation in terms of the
Alexander-Anderson model~\cite{AlexanderAnderson} see
\cite{lounis}), and stronger than the adatom-substrate
interaction, the Ferri solution is expected to prevail. Indeed,
Cr dimers on Ni(111) as on Ni(001) are characterized by a collinear
Ferri coupling as a ground state. The difference is, however, that no
non-collinear solution was found on Ni(111), as opposed to
Ni(001)~\cite{lounis}. This is understandable, because the
non-collinear state in the dimer on Ni(001) arises from the
competition between the intra-dimer Cr-Cr antiferromagnetic
interaction and the Cr-Ni antiferromagnetic interaction. On the
Ni(111) surface, the coordination to the Ni substrate is lower,
therefore the interaction with the substrate is insufficient to
overcome the Cr-Cr interaction. In fact the
Ferri total energy is 317.32 meV/adatom lower than the AF one and
352.54 meV/adatom lower than the FM one.

\begin{table}
\begin{center}
\begin{tabular}{lccc}
\hline
    & & Cr$_2$ & Mn$_2$ \\ 
\hline 
AF  & moments  &($-3.47, -3.47$)&($-4.02, -4.02$)\\         
  & $E_{\mathrm{AF}} -E_{\mathrm{Ferri}}$&317.32 &243.2 \\ 
\hline 
Ferri& moments &($-3.30, 3.31$)&($-3.97, 3.85$)\\
\hline 
FM&moments&  ($3.38, 3.38$) &($3.98, 3.98$)\\
  & $E_{\mathrm{FM}} -E_{\mathrm{Ferri}}$&352.54 &76.31 \\
\hline
\end{tabular}
\caption{Atomic spin moments (in $\mu_B$) and energy differences (in
meV) of the adatom dimers on Ni(111) in the collinear
configurations. A minus sign of the collinear moments indicates an
antiparallel orientation with respect to the substrate magnetization.
}
\label{dimermomNi111}
\end{center}
\end{table}

Similar trends are found for the Mn dimers on Ni(111): The Ferri
solution is the most stable collinear solution. However, in addition a
non-collinear solution is found, which is only slightly higher, {\it
i.e.} by 4.44 meV/adatom than the Ferri solution. Note that on the
Ni(001) surface\cite{lounis}, this type of dimer state, shown in
fig.~\ref{dimer-Ni111}(d), represents the ground state, which is,
however, not the case for the dimer on (111) prefering the Ferri
configuration. In the
non-collinear configuration (fig.~\ref{dimer-Ni111}(d)), both adatom
moments (3.90 $\mu_B$), while aligned antiferromagnetically with
respect to each other, are slightly tilted in the direction of the substrate
magnetization with a rotation angle of 
$\theta=79^{\circ}$ (instead of 90$^{\circ}$). The energy
differences between the Ferri with the other local minima AF and FM
are respectively 243.2 meV/adatom and 76.31 meV/adatom.  The magnetic
moments and energy differences are given in
Table~\ref{dimermomNi111}. Note that the local Cr and Mn moments are
considerably higher than the corresponding local moments in dimers on
Ni(001)\cite{lounis}, again a result of the reduced coordination
number.
\begin{table}
\begin{tabular}{rcccc}
\hline
 Dimer type:        & Cr$_2$  & Cr$_2$  & Mn$_2$  & Mn$_2$ \\
 Substrate: & \ Ni(111) \  & \ Ni(001) \  & \ Ni(111) \ & \  Ni(001) \ \\
\hline
$E_{\FM-\Ferri}$  & 353 & 451 & \phantom{0}76  &  \phantom{0}65 \\
\hline
$E_{\AF-\Ferri}$  & 317 & 433 & 243 & 496 \\
\hline
\end{tabular}
\caption{\label{table:dimer_energy} Dimer energies (in meV/adatom) 
in the FM, AF, and Ferri configurations of Cr and Mn 
dimers on Ni(111). Results of the same dimers on Ni(001), 
taken from Ref.~\onlinecite{lounis}, are also shown for comparison.}
\end{table}

As we have discussed in Ref.~\onlinecite{lounis}, the different
magnetic configurations of the Ni substrate atoms cannot be described
well by the Heisenberg model, since the moments of the atoms adjacent
to the adatoms are strongly reduced. Such longitudinal moment
relaxations cannot be described by this model. 

\section{Adatom Trimers}
\begin{figure}
\includegraphics*[width=\linewidth]{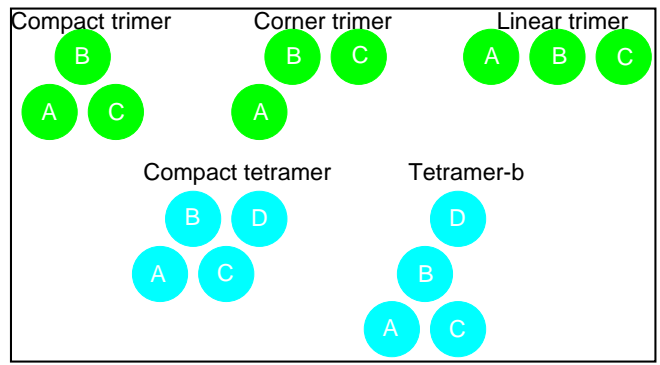}
\caption{Different geometrical configurations 
considered for trimers and tetramers 
at the surface of Ni(111).
}
\label{geometry-Ni111}
\end{figure}
Trimers in equilateral triangle geometry are, in
the presence of antiferromagnetic interactions, prototypes for
non-collinear magnetism, with the magnetic moments of the three atoms
having an angle of 120$^{\circ}$ to each other\cite{stocks}. This
\emph{120$^{\circ}$-configuration} is a well-known consequence of the
magnetic frustration in such triangular systems. Examples are compact Cr
or Mn trimers on (111) surfaces of noble metals (see for instance a
recent calculation of Mn$_3$ on Cu(111) reported in
Ref.~\onlinecite{bergman}). In our case, the 120$^{\circ}$ state is
perturbed by the exchange interaction with the substrate, 
and therefore the
magnetic configuration is expected to be more complicated.

Let us start with a Cr dimer (Mn dimer) that we approach by a single
Cr adatom (Mn adatom). As shown in Fig.\ref{geometry-Ni111}, three
different types of trimers can be formed: i) the compact trimer with
an equilateral shape, ii) the corner trimer with an isosceles shape
and iii) the linear trimer. The adatoms are named A, B and C.

\begin{figure}
\begin{center}
{(a)}
\includegraphics*[width=0.4\linewidth]{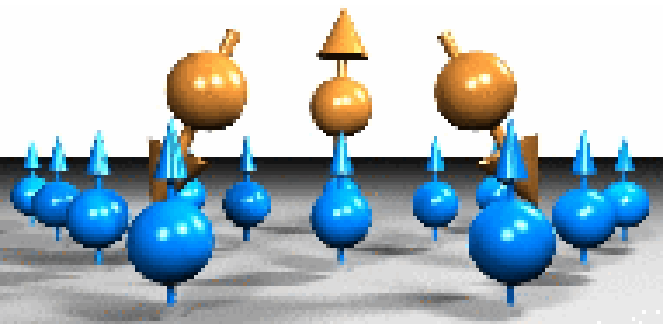}
\hspace{-0.2cm}
{(b)}
\includegraphics*[width=0.4\linewidth]{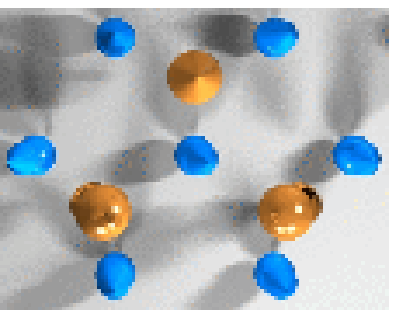}
\\
{(c)}
\includegraphics*[width=0.4\linewidth]{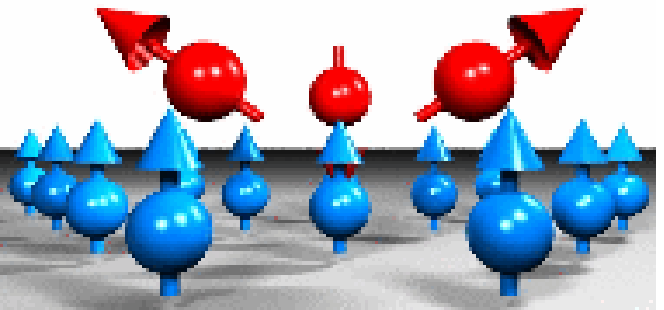}
\hspace{-0.2cm}
{(d)}
\includegraphics*[width=0.4\linewidth]{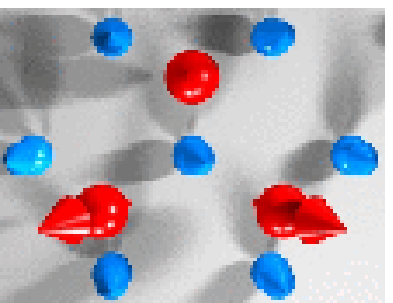}
\\
{(e)}
\includegraphics*[width=0.4\linewidth]{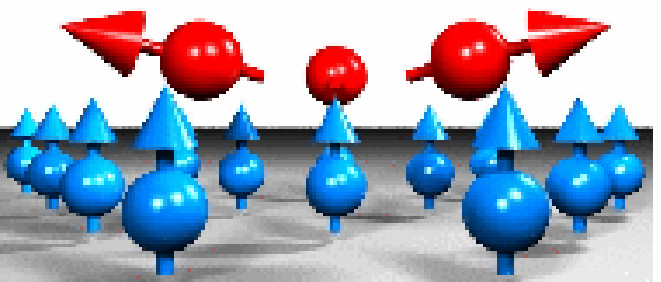}
\hspace{-0.2cm}
{(f)}
\includegraphics*[width=0.4\linewidth]{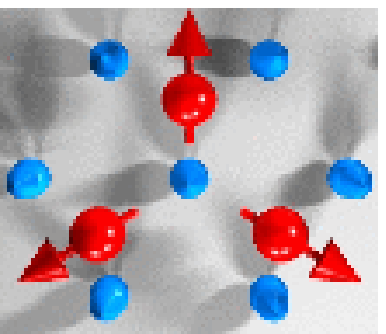}
\\
{(g)}
\includegraphics*[width=0.4\linewidth]{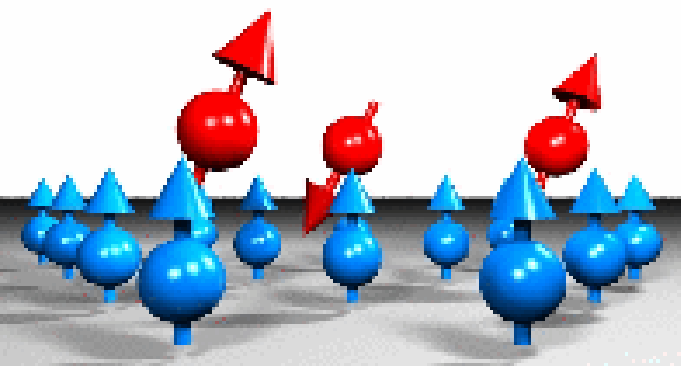}
\hspace{-0.2cm}
{(h)}
\includegraphics*[width=0.4\linewidth]{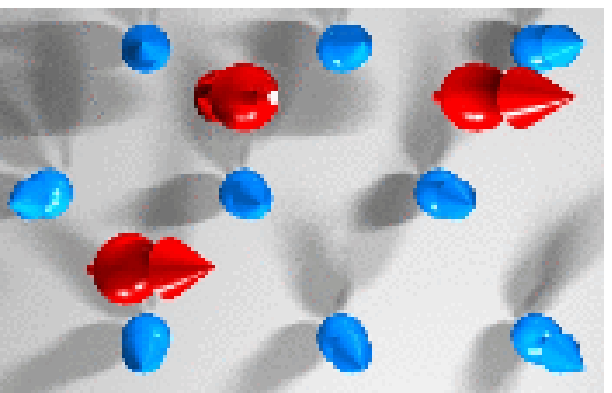}
\caption{ Side view (a) and top view (b) are shown for the most stable
configuration of Cr compact trimer on Ni(111) (in blue). (c) and (c) represent the 
side view and top  view of the ground state (NCOL2) of Mn compact trimer on Ni(111) 
while (d) and (e) depict an almost 
degenerate state (NCOL3) of the same Mn trimer. Finally, the
side view (g) and top view (h) are shown for the most stable
configuration of Mn (in red) corner trimer on Ni(111).
(see text for more details).}
\label{trimer-cr-ni111}
\end{center}
\end{figure}

When the distance between the ad-dimer and the single adatom is large
enough that their magnetic interaction is weak (second-neighboring
positions\cite{lounis}), the total moment is $-3.78\ \mu_B$ for the Cr
case and 4.05 $\mu_B$ for the Mn one.  Let us move the single adatom
close to the dimer and form a compact trimer. The distance between the
three adatoms is the same, meaning that this is a prototype geometry
which leads for a trimer in free space to a 120$^{\circ}$ rotation
angle between the magnetic moments. This is attested for the Cr case
for which we had difficulties finding a collinear solution. Our
striking result, as depicted in fig.~\ref{trimer-cr-ni111}a-b, is that
the non-collinear 120$^{\circ}$ configuration is conserved with a
slight modification.  Indeed, our self-consistent ($\theta$,
$\phi)$-angles are ($2^{\circ}, 0^{\circ}$) for adatom B and
($126^{\circ}, 0^{\circ}$) for adatom A and ($122^{\circ},
180^{\circ}$) for adatom C. The angle between B and A is equal to the
angle between B and C (124$^{\circ}$) while the angle between A and C
is 112$^{\circ}$. The small variation from the prototypical
120$^\circ$ configuration is due to the additional exchange
interaction with Ni atoms of the surface. Let us suppose that we start
with a 120$^{\circ}$ configuration of a compact Cr trimer, neglecting
at first the exchange interaction with the substrate. This gives an
infinite number of degenerate configurations being distinguished by an
arbitrary rotation of all moments in spin space. This degeneracy is
(partly) removed by coupling to the substrate atoms, the moments of
which are fixed by anisotropy, {\it e.g.} in [111] direction. Since 
the adatom-substrate interaction is AF, the
moments of two adatoms rotate so that they are partly oriented
opposite to the substrate magnetization, while the moment of the third
adatom rotates to the opposite direction, driven by the AF interaction
to its Cr neighbors.  The coupling with the substrate leads thus to a
deviation from the prototype 120$^\circ$ state, with an additional
rotation of $2^{\circ}$ for the FM adatom and of 4$^{\circ}$ for the
two other adatoms.  The ferromagnetic Cr carries a moment of 2.94
$\mu_B$, smaller than the neighboring moments (3.31 $\mu_B$). Hence,
the total magnetic moment of all the adatoms is -0.76 $\mu_B$. Note
the huge jump of the total magnetic moment (80\%) from -3.78 $\mu_B$,
which is the initial non-interacting dimer-adatom total moment.

For the compact Mn trimer, three non-collinear configurations were
obtained: As in the case of the compact Cr trimer, the free Mn trimer
must be in a 120$^{\circ}$ configuration. Nevertheless, the magnetism
of the substrate changes this coupling taking into account the single
adatom behavior: Mn adatoms prefer a FM coupling to the substrate and
an AF coupling with their neighboring Mn adatom. The first
non-collinear magnetic configuration (NCOL1) is similar to the Cr one
(fig.~\ref{trimer-cr-ni111}a-b), {\it i.e.} adatom B couples FM (3.61
$\mu_B$) with the substrate moments while adatom A (3.67 $\mu_B$) and
C (3.67 $\mu_B$) are rotated into the opposite direction with an angle
of 114$^{\circ}$ between B and A and between B and C 
(see Table~\ref{table:Mn-trimer-moment}). The second
non-collinear configuration (NCOL2) has the opposite magnetic picture
(fig.~\ref{trimer-cr-ni111}c-d) as compared to compact Cr trimer {\it
i.e.}: there is a rotation of the moment of atom B (3.70 $\mu_B$) to
an almost AF coupling ($\theta$ = 179$^{\circ}$, $\phi$ = 0$^{\circ}$)
forced by the two other adatoms (A and C) which tend to couple FM to
the substrate by experiencing an additional rotation angle of
10$^{\circ}$ towards a FM coupling {\it i.e.} atom A carries a moment
of 3.62 $\mu_B$ rotated by ($\theta$ = 49$^{\circ}$, $\phi$ =
0$^{\circ}$) instead of ($\theta$ = 60$^{\circ}$, $\phi$ =
0$^{\circ}$) which characterize the free Mn trimer. On the other hand,
atom B has a moment of 3.62 $\mu_B$ rotated by ($\theta$ =
51$^{\circ}$, $\phi$ = 180$^{\circ}$).  The angle between B and A is
equal to the angle between B and C (50$^{\circ}$) while the angle
between A and C is 260$^{\circ}$. In the third magnetic configuration
(NCOL3) the three moments (3.65 $\mu_B$) are almost in-plane and
perpendicular to the substrate magnetization (see
fig.~\ref{trimer-cr-ni111}e-f). They are also slightly tilted in the
direction of the substrate magnetization ($\theta = 86^{\circ}$) due
to the weak FM interaction with the Ni surface atoms. Within this
configuration, the 120$^{\circ}$ angle between the adatoms is kept.
Total energy calculations show that the NCOL2 configuration is the
ground state which is almost degenerate with NCOL1 and NCOL3 ($\Delta
\mathrm{E}_{\mathrm{NCOL1}-\mathrm{NCOL2}} = 1.27$ meV/adatom and
$\Delta \mathrm{E}_{\mathrm{NCOL3}-\mathrm{NCOL2}} = 5.61$
meV/adatom). Thus already at low temperatures trimers might be 
found in all three configurations; in fact the spin arrangement 
might fluctuate between these three 120$^{\circ}$ configurations or even at 
low temperatures, between the three degenerate configurations of the NCOL1 or the 
NCOL2 state. Compared to the collinear 
state energy of the compact trimer, 
the NCOL2 energy
is lower by 138.23 meV/adatom. This very high energy difference 
is due to frustration, even higher than breaking a bond as shown 
in the next paragraphs. Contrary to this, the corner trimer shows no 
frustration which leads to a collinear ground state. For instance, the total moment 
of the non-collinear compact trimer 
(0.95$\mu_B$) experiences a decrease of 76\% compared to the obtained
value for the non-interacting dimer-adatom configuration. 

\begin{table}
\begin{tabular}{lcccc}
\hline
Noncol. config. & Adatom & Moment ($\mu_B$) & $\theta$(deg) & $\phi$(deg) \\
\hline
 & A                                &3.67    & 115   & 0 \\  
 NCOL1 &B                           &3.61    & 1     & 0 \\
 & C                                &3.67    & 113   & 180\\
\hline
 & A                                &3.62    & 49    & 0 \\  
 NCOL2 &B                           &3.70    & 179   & 0 \\
 & C                                &3.62    & 50    & 180\\
\hline
 & A                                &3.65    & 86    & 120 \\  
 NCOL3 &B                           &3.65    & 86    & 0 \\
 & C                                &3.65    & 86    & 120\\
\hline
\end{tabular}
\caption{\label{table:Mn-trimer-moment}Atomic moments 
and rotation angles of the magnetic moments of Mn 
adatoms forming a compact trimer on Ni(111) surface. For 
the adatom notation see fig.~\ref{geometry-Ni111}.}
\end{table}

The next step is to move the additional adatom C and increase its
distance with respect to A in order to reshape the trimer into an
isosceles triangle (what we call ``corner trimer'' in
fig.~\ref{geometry-Ni111}, with one angle of 120$^{\circ}$ and two of
30$^{\circ}$). By doing this, the trimer loses the frustration and is
characterized, thus, by a collinear ferrimagnetic ground state: the
moments of adatoms A and C are AF oriented to the substrate (following
the AF Ni-Cr exchange), while the moment of the central adatom B is FM
oriented to the substrate, following the AF Cr-Cr coupling to its two
neighbors. The magnetic moments do not change much compared to the
compact trimer. The central adatom B carries a moment of 2.94 $\mu_B$
while the two others have a slightly more sizeable moment of $-3.32\
\mu_B$. Thus, the total magnetic moment ($-3.70\ \mu_B$) increases to
a value close to the one obtained for non-interacting dimer-adatom
system.

While the non-collinear state is lost for the corner Cr trimer, it is
present for the corner Mn trimer as a local minimum with a tiny energy
difference of 4.82 meV/adatom higher than the Ferri ground state.
This value is equivalent to a temperature of $\sim$ 56~K, meaning that
at room temperature both configurations co-exist. Here Ferri means
that the central adatom B is AF oriented to the substrate with a
magnetic moment of 3.71 $\mu_B$, forced by its two FM companions A and
C (moment of 3.83 $\mu_B$) which have only one first neighboring
adatom and are less constrained. The total moment of the adcluster is
also high ($3.95 \mu_B$) compared to the compact trimer value,
reaching the value of the non-interacting system (with the third atom
of the trimer far away from the other two).

The Ferri solution is just an extrapolation of the non-collinear
solution shown in fig.~\ref{trimer-cr-ni111}g-h (with magnetic moments
similar to the collinear ones) in which the central adatom B (3.70
$\mu_B$) tends to orient its moment also AF to the substrate ($\theta$
= 152$^{\circ}$, $\phi$ = 0$^{\circ}$) and the two other adatoms with
moments of 3.83 $\mu_B$ tend to couple FM to the surface magnetization
with the same angles ($\theta$ = 23$^{\circ}$, $\phi$ =
180$^{\circ}$).  It is important to point out that the AF coupling
between these two latter adatoms is lost by increasing the distance
between them. Indeed, one sees in fig.~\ref{trimer-cr-ni111}g-h that
the two moments are parallel. The total magnetic moment is also high
and is equal to 3.78 $\mu_B$.

Let us move furthermore adatom C in order to form a linear
trimer. For the Cr case, there is no non-collinear magnetism. The
already stable Ferri solution for the corner trimer is comforted. The
moments of the adatoms A and C are a bit higher than what obtained so
far for the other structural configuration, {\it i.e.}, adatoms A
and C have a moment of $-3.40\ \mu_B$ while the central moment is equal
to 2.97 $\mu_B$: the coupling between A and C is now indirect (through
the central adatom). The total magnetic moment is also high (-3.83
$\mu_B$).

Concerning the Mn case for a chain, a non-collinear configuration was
obtained as a local minimum with a small energy difference 
compared to the ground state which is the collinear
Ferri solution (8.50 meV/adatom $\sim$ 98.64 K). The magnetic 
moments do not change a lot compared to the
values obtained for the corner trimer. The central adatom B carries a
moment of $-3.78\ \mu_B$ while the A and C have a higher moment of 3.84
$\mu_B$. In the non-collinear solution, the central Mn adatom with a
moment of 3.76 $\mu_B$ , as seen previously, tends to couple AF with
($\theta$ = 142$^{\circ}$, $\phi$ = 0$^{\circ}$) and the A and C with
a similar moment of 3.85 $\mu_B$ tend to couple FM to the substrate
($\theta$ = 28$^{\circ}$, $\phi$ = 180$^{\circ}$). The total moment is
high for both magnetic configurations. For the Ferri solution, it
reaches 3.90 $\mu_B$, while for the non-collinear solution the total
moment value is smaller (3.84 $\mu_B$).

It is interesting to compare the total energies of the three trimers
we investigated. The compact trimer has more first neighboring bonds
and is expected to be the most stable trimer. The energy differences
confirm this statement. Indeed the total energy of the Cr compact
trimer is 119 meV/adatom lower than the total energy of the corner
trimer and 198.16 meV/adatom lower than the total energy of the linear
trimer. Similarly, the Mn compact trimer has a lower energy of 53
meV/adatom compared to the corner trimer and a lower energy of 100
meV/adatom than the linear trimer. 

By summarizing the results for the total moments of the adatoms, we
see that the non-interacting cluster consisting of a single adatom and
an adatom dimer has a high moment. This large total moment also
survives for linear and corner trimers. However the most stable
compact trimer has a low moment, 0.95 $\mu_B$ in the case of Mn and
-0.76 $\mu_B$ for Cr.

\section{Adatom Tetramers}
\begin{figure}
\begin{center}
{(a)}
\includegraphics*[width=0.7\linewidth]{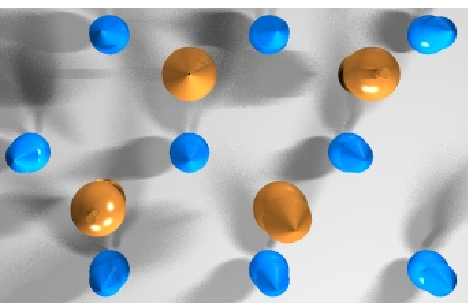}\\
{(b)}
\includegraphics*[width=0.7\linewidth]{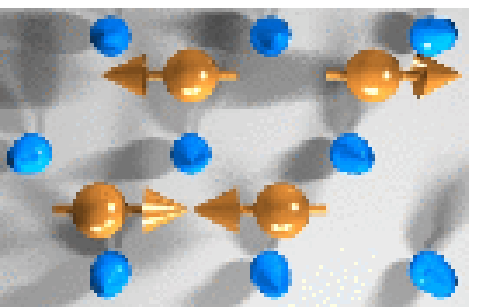}
\\
{(c)}
\includegraphics*[width=0.7\linewidth]{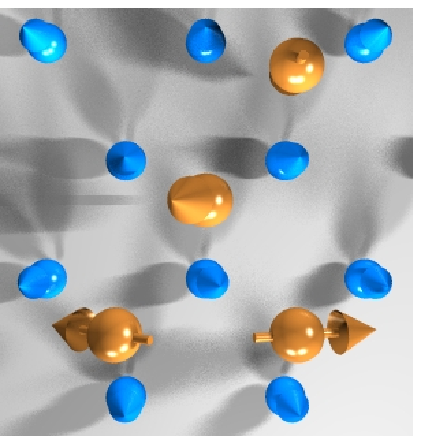}
\end{center}
\caption{Top view of the collinear most stable solution (a) and the non-collinear 
metastable configuration (b) of compact Cr~tetramer 
on Ni(111). In (c) is depicted 
The Cr~tetramer-b magnetic ground state on 
Ni(111), which basically consists of the non-collinear 
trimer state of fig.\ref{trimer-cr-ni111}a coupled antiferromagnetically to the 
fourth adatom.}
\label{tetramer-cr-ni111}
\end{figure}
We consider two types of tetramers, formed by adding a Cr or Mn adatom
(atom D in fig.~\ref{geometry-Ni111}) to the compact trimer.  We begin
with the compact tetramer (see fig.~\ref{geometry-Ni111} and
fig.~\ref{tetramer-cr-ni111}a-b). For both elements Cr and Mn, the
Ferri solution is the ground state
(fig.~\ref{tetramer-cr-ni111}a). For Cr (Mn) compact tetramer, the A
and D adatoms are FM oriented to the surface atoms with a moment of
2.31 $\mu_B$ (3.60 $\mu_B$) while B and C are AF oriented to the
substrate with a moment of 2.87 $\mu_B$ (3.43 $\mu_B$). This gives a
magnetic configuration with a low total magnetic moments of $-1.12\
\mu_B$ for Cr tetramer and 0.34 $\mu_B$ for Mn tetramer. The Cr
tetramer, in particular, shows also a non-collinear configuration
(fig.~\ref{tetramer-cr-ni111}b) as a local minimum which has, however,
a slightly higher energy of $\Delta
\mathrm{E}_{\mathrm{NCOL}-\Ferri} = 1$ meV/adatom.  
Within this configuration the AF coupling between the
adatoms is observed. However, the four moments are almost in-plane
perpendicular to the substrate magnetization. The tilting is small
($\theta$ = 93$^{\circ}$) due to the weak AF coupling with the
substrate.

An additional manipulation consists in moving 
the adatom D and forming 
a tetramer-b (fig.~\ref{tetramer-cr-ni111}c). For such a structure, the
collinear solution for the Cr tetramer is only a local minimum. In
this structure, atom D has less neighboring adatoms compared to A, B,
and C. In the non-collinear solution which is the magnetic ground
state, the moment of adatom D (3.34 $\mu_B$) is almost AF oriented to
the substrate ($\theta$ = 178$^{\circ}$, $\phi$ = 0$^{\circ}$). The
remaining adatoms form a compact trimer in which the closest adatom to
D, {\it i.e.} B, tends to orient its moment FM (2.45 $\mu_B$) to the
substrate ($\theta$ = 19$^{\circ}$, $\phi$ = 0$^{\circ}$) while the
moments of A (2.90 $\mu_B$) and C (2.80 $\mu_B$) tend to be oriented
AF ($\theta_{\mathrm{A}}$ = 124$^{\circ}$, $\phi_{\mathrm{A}}$ =
0$^{\circ}$) and ($\theta_{\mathrm{C}}$ = 107$^{\circ}$,
$\phi_{\mathrm{C}}$ = 180$^{\circ}$).  In the (metastable) collinear
solution for this tetramer, the moment of adatom B is oriented FM to
the substrate while the moments of all remaining adatoms are oriented
AF to the surface atoms.  Here, the total magnetic moment has a high
value of $-3.46\ \mu_B$.  The total energy difference between the two
configurations is equal to 49.32 meV/adatom. Compared to the total
energy of the compact tetramer, our calculations indicate that the
tetramer-b has a higher energy (108.72 meV/adatom).  

Let us now turn to the case of the Mn tetramer-b. Also here, the
non-collinear solution is the ground state while the collinear one is
a local minimum. The energy difference between the two solutions is
very small (2.82 meV/adatom).  The moments are now rotated to the
opposite direction compared to the Cr case, in order to fulfill the
magnetic tendency of the single Mn adatom which is FM to the
substrate. The Mn atom with less neighboring adatoms, {\it i.e.} D,
has a moment of 3.84 $\mu_B$ rotated by ($\theta$ = 27$^{\circ}$,
$\phi$ = 0$^{\circ}$), while its closest neighbor, the atom B with a
moment of 3.44 $\mu_B$, is forced by the neighboring companions to
couple AF ($\theta$ = 140$^{\circ}$, $\phi$ = 180$^{\circ}$). The
adatoms A and C with similar magnetic moments (3.63 $\mu_B$) tend to
couple FM with the following angles: ($\theta$ = 81$^{\circ}$, $\phi$
= 0$^{\circ}$) and ($\theta$ = 34$^{\circ}$, $\phi$ = 0$^{\circ}$). As
in the case of Cr tetramer-b, the converged collinear solution is just
the extreme extension of the non-collinear one: The ``central'' adatom
of the tetramer is forced by its FM Mn neighboring atoms to coupled AF
to the substrate. The magnetic regime is similar to the one of Cr
tetramer-b, {\it i.e.} high, with a total magnetic moment of 4.37
$\mu_B$.

As expected, the most stable tetramer is the compact one, with an
energy of 52.26 meV/adatom lower than tetramer b.

\section{Summary}

As a summary, we have investigated the complex magnetism of small Cr
and Mn ad-clusters on Ni(111). This is a prototype system where 
two types of magnetic frustration occur: (i) 
frustration within the ad-cluster and (ii) frustration arizing from 
antiferromagnetic coupling between the adatoms in the cluster and competing 
magnetic interaction between the adclusters and the surface atoms. 
While the resulting collinear and non-collinear structures are very 
complex, a unifying feature is that all compact structures (dimers, trimers and 
tetramers) have very small total moments, as a result of the strong 
antiferromagnetic coupling between the cluster atoms leading to 
a nearly complete compensation of the local moments, while the more open 
structures, like the corner and linear trimers and the tetramer b, have 
rather large total moments of about 4 $\mu_B$. Since the transtion 
between a compact and an open structure requires to move one adatom by just 
one atomic step, we might consider this motion as a magnetic switch, which 
via the local magnetic exchange field of the single adatom allows to switch 
the total moment on and off, and which therefore might be of interest for magnetic 
storage. Thus magnetic frustration might be useful for future nanosize information 
storage.

\section*{acknowledgments}
This work was financed by the Priority Program ``Clusters in Contact
with Surfaces'' (SPP 1153) of the Deutsche Forschungsgemeinschaft.

\end{document}